\documentclass[conference]{IEEEtran}
\IEEEoverridecommandlockouts
\usepackage{cite}
\usepackage{amsmath,amssymb,amsfonts}
\usepackage{algorithmic}
\usepackage{graphicx}
\usepackage{textcomp}
\usepackage{xcolor}
\usepackage{listings}
\usepackage{listings-owl}
\usepackage{hyperref}
\def\BibTeX{{\rm B\kern-.05em{\sc i\kern-.025em b}\kern-.08em
    T\kern-.1667em\lower.7ex\hbox{E}\kern-.125emX}}
\begin{document}

\title{Managing Differentiated Secure Connectivity using Intents}

\author{\IEEEauthorblockN{Loay Abdelrazek}
\IEEEauthorblockA{\textit{Standards \& Technology} \\
\textit{Ericsson}\\
Sweden\\
first.last@ericsson.com}
\and
\IEEEauthorblockN{Filippo Rebecchi}
\IEEEauthorblockA{\textit{Standards \& Technology} \\
\textit{Ericsson}\\
France \\
first.last@ericsson.com}
}

\maketitle

\begin{abstract}
Mobile networks in the 5G and 6G era require to rethink how to manage security due to the introduction of new services, use cases, each with its own security requirements, while simultaneously expanding the threat landscape. Although automation has emerged as a key enabler to address complexity in networks, existing approaches lack the expressiveness to define and enforce complex, goal-driven, and measurable security requirements. 
In this paper, we propose the concept of differentiated security levels and leveraging intents as a management framework. We discuss the requirements and enablers to extend the currently defined intent-based management frameworks to pave the path for intent-based security management in mobile networks. Our approach formalizes both functional and non-functional security requirements and demonstrates how these can be expressed and modeled using an extended TM Forum (TMF) intent security ontology. We further discuss the required standardization steps to achieve intent-based security management. Our work aims at advance security automation, improve adaptability, and strengthen the resilience and security posture of the next-generation mobile networks
\end{abstract}


\section{Introduction}
Mobile networks are the digital backbone of today's societies and economies, and ensuring their security and trustworthiness is a priority \cite{europarl5g,nis2}. In this context, security management refers to the continuous identification, detection and mitigation of threats, and reporting of the security status of a Network Function (NF), infrastructure, and services. In the different domains of a mobile network, for example Radio Access Network (RAN), Core Network (CN) and Transport Network (TN), this occurs both during deployment and operation, with continuous risk and trust monitoring to support risk-based decision making. 

The evolution toward 5G and 6G has increased network complexity, due to a growing number of subsystems, technologies, and services that must be supported \cite{cagenius20236g}\cite{li2021complexity}. At the same time, mobile networks face of an expanding threat landscape, including, among many others, supply chain attacks, advanced persistent threats (APTs), and protocol-specific exploits \cite{gsmasec24}. Secure and resilient operations rely on proper configuration of a multitude of settings, policies, and service parameters.
Automation is increasingly adopted to address growing system complexity, reduce operational costs and errors, and alleviate the workload of operation teams. Technologies such as software-defined networking (SDN), network function virtualization (NFV), and advanced orchestration platforms accelerate service delivery. 

In security, automation facilitates zero-touch security management, allowing security teams to define high-level objectives and constraints, while automation logic interprets, translates, actuates, and enforces them without manual intervention. Automated security management also helps mitigating workforce and skills shortages by minimizing repetitive manual tasks and streamlining risks identification and threat detection. The main objective is to reduce both mean-time-to-detect (MTTD) and mean-time-to-respond (MTTR) metrics, regardless of whether responses are carried out by humans (in an open-control loop), or autonomously by the system (in a closed control loop).

Despite these advances, current automation platforms lack the expressiveness to define complex and goal-driven security requirements. Intent-based Management (IbM) offers a promising approach: by expressing desired outcomes as high-level declarative intents, IbM can manage the security complexity and adapt to the different requirements
~\cite{IG1253}. Leveraging intents will allow NFs to adopt self-protection capabilities as of an autonomous network. The main aim is to support the system to autonomously monitor, analyze, and respond to threats within a closed control loop.
For example, a service provider can use security intents to simplify the deployment and design phases, ensure regulatory compliance, as well as proactively protect NFs against attacks and possible threats. While this vision is compelling, there is still no standardized method for formalizing security intents and translating them into measurable service characteristics to achieve differentiated security capabilities. Hence, the correct enforcement of security requirements remains largely manual, hindering automation, and preventing networks' ability to fully adapt to an evolving threat landscape.  

In this paper, our contributions are as follows:
\begin{itemize}
    \item We propose the concept of differentiated security levels and manage them in complex mobile networks via IbM.
    \item We formalize both functional and non-functional requirements for security, that can be measurable both qualitatively and quantitatively and enable true goal-driven security management. 
    \item We demonstrate such security requirements can be expressed via intents, by extending the existing intent model from the TM Forum (TMF).
    \item We discuss the required extensions to current standards to support intent-based security management.
\end{itemize}


The rest of the paper is organized as follows: Section~\ref{background} provides background on security requirements used and IbM. Section~\ref{related_work} reviews the state-of-the-art in mobile network security management focusing on work proposed Security Service Level Agreements and Zero-touch management. Section~\ref{diff_sec_req} introduces the proposed differentiated security requirements. Section~\ref{sec_req_intents} provides examples of expressing security expectations in an intent model. Section~\ref{standards} discusses initial standardization directions for intent-based security management. Finally, Section~\ref{conclusion} concludes the paper.

\section{Background}\label{background}

\subsection{Security Requirements}\label{sub_securityrequirements}
One key contribution of this paper is to provide a representation of the security requirements in a standardized way to enable the adaptation of security. To align on what a security requirements is, we adopt the definition from NIST Special Publication 800-160, namely: \textit{the statement of needed security functionality that ensures one of many different security properties of system is being satisfied}~\cite{nist800-160}. 

Security requirements can be derived from industry standards, applicable regulations, but also from the history of past vulnerabilities. A security requirement defines new capability (e.g, protection, detection, mitigation capability) or additions to existing capability to solve a specific security problem or eliminate a potential vulnerability. Additionally, system security requirements define the protection capabilities provided by the system, the performance and behavioral characteristics exhibited by the system, and the evidence used to determine that the system security requirements have been satisfied. Generally, security requirements can be categorized as:
\begin{enumerate}
    \item \textbf{Functional requirements} define the security functions and capabilities that a system must have to fulfill different security properties. Key properties to be achieved: confidentiality, isolation, integrity, privacy, authentication, detection, etc. In the following, we will refer to this category as \textbf{Protection Requirements}.
    \item \textbf{Non-functional requirements} define how security controls must perform or behave. Key properties to be achieved: robustness, control performance, detection time, mitigation impact, etc. In the following, we will refer to this category as \textbf{Security Performance Requirements}. 
\end{enumerate}

\subsection{Intent-based Management}
The telecommunication industry is increasingly adopting management automation to reduce complexity and Operating Expenses (OPEX), while better adapting to the business objectives and customer expectations. This transformation also includes security management. 

Intent-based systems provide a promising framework for embedding autonomous capabilities within security management. \textbf{A fundamental requirement for intent-driven operations is the ability to formalize requirements as explicit goals}. Both TMF and~3GPP define \textit{intents} as  declarative statements that specifies “what” outcomes are expected, rather than “how” they should be achieved~\cite{IG1253,3ggp28312}. An intent provides the (sub-)system with the formal specification of expectations (these include requirements, goals and constraints). In turn, each (sub-)system leverages its knowledge of the overall system state to autonomously select and enforce the best solutions to fulfill these requirements, subject to the provided constraints.

Therefore, a homogeneous definition of security intents is a necessary building block for standardizing differentiated security capabilities, enabling their adaptation to the specific needs and risk profiles of slices, customers, and use cases. 

\section{Related Work}\label{related_work}
There is work in the domain of security management and automation that explored intent-based management and its usage for security use cases to achieve Zero-touch security networks in 5G and beyond. To the best of our knowledge, the papers mostly investigated the architectural aspects of intent-based security management specifically on how to apply the ETSI Zero-touch Service Management Architecture \cite{etsizsm} to the security domain. However, the work did not investigate the possible security goals and expectations and how they can be expressed in an intent model, neither possible standardization requirements to ensure the operability in networks.
The work done in \cite{ortiz2020inspire} highlights a security management platform developed as part of the INSPIRE-5Gplus project \cite{inspire5g}, where they leveraged technologies like machine learning (ML) to enable a closed-loop security automation and end-to-end security management following the zero-touch paradigm. One of the key aspects of the platform is the ability to manage security using Security Service Level Agreements (SSLA). The defined SSLA does not define an expectation or a goal that the network to fulfill, it rather defines the security capability to be part of the network (e.g, confidentiality, integrity, detection of attacks, etc.. ) that is needed part of the network. In this paper, we rather argue that a security capability is a solution part of a solution space that a system can automatically analyze and select the suitable controls to achieve a quantified goal. Additionally, adopting capabilities as requirements limits the opportunity that different assets of the network can have different requirements, for example a network function that has less sensitive data would have less strict protection level requirement, than a network function that is on the edge of the network. As a result, the network function utilize a 256-bit encryption algorithm on the edge network function but a 128-bit encryption algorithm on the network function with less stricter requirements. Consequently, different controls and different solution spaces lead to more flexibility on how to fulfill a requirement and balances security requirements with its impact on the network. Moreover, the aforementioned work did not consider expressing the requirements as intent models, which is the main management paradigm discussed today to manage future networks. 

Additionally, the work in \cite{chollon2022etsi} is also part of \cite{inspire5g} discusses the aspect of managing security through SSLA. However, they define SSLA to be expressed as policies and not intents. Where policies are formulated by specifying the security property to be achieved (e.g, Confidentiality) along with the solution that is manifested in the security capability (e.g, AES-CBC encryption algorithm with 128-bits key length). Although this is the current known paradigm to manage security, it becomes a burden in future network where there are different solutions, different services with their own Key Performance Indicators (KPIs) to be maintained, and it is expected from a human operator to configure the security policies. This approach consumes time to configure and does not scale. However, if less human intervention is allowed where the system is able to select the suitable solutions based on a goal, and human operators are always able to intervene to modify decisions, this approach will reduce the time to implement new services in the network, and enable to manage security simply and efficiently.

On the other hand, in \cite{wang2023intent} the authors discussed utilizing intent-based management to provision secure slices based on security requirements, for the objective of simplifying security controls management and minimize the gap between the security requirements and the security capabilities to be provisioned. Even though they discussed life cycle management of intent, decomposition and translating into security configurations, what is closest to our work is the authors attempt to define different security capabilities for provisioning a secure network slice. In our work we argue that capabilities as defined in \cite{3ggp28312} are the solution space or the set of actions to be actuated. The actuated actions should fulfill the requirement, thus the goals expressed in the intent should not be "how" to achieve a goal it is rather expressing "what" is the goal. Therefore, providing a set of functional capabilities and grouping them together in a level is counter-intuitive as the automation logic in a autonomous system is responsible on analyzing the available solutions and plan the most suitable capabilities and infer its configuration to fulfill the requirement and in the context of any available constraints (e.g, conditions to maintain other KPIs at a certain level). Moreover, they provide a mapping between the different levels of protection, which is not defined what is a protection level, and if that can be a quantified requirement. In addition, they map the different qualitative levels to performance metrics (e.g, latency) which was yet again defined as qualitative impact level (High, Medium, Low) which does not provide any concrete indication of the magnitude and impact and , as this could be a useful input to conflict management part of an IbM system. Additionally, the type of requirements were defined are functional requirements which is a good starting point, but we argue that it is important to track the performance of security and allow automation loops within the network functions to take decisions to continuously adapt the controls in order to proactively identify, protect, detect and mitigate attacks.

\section{Differentiated Security Requirements}~\label{diff_sec_req}
Defining the security requirements in a goal-driven and modular method enables the ability to program different security capabilities end-to-end in a mobile network. The differentiation in the security capabilities is not only on the type of control (i.e. firewall, IPsec, Radio Encryption) but also how the capabilities are configured to meet the different security level requirements in addition to balance with performance impact and cost on resources (e.g, cpu utilization, bandwidth overhead, latency,etc..). 

The ability to manage a network with different security requirements not only in different parts of the network but also on different resources within the same network function, transforms the security domain from a binary solution domain that can either be enabled or disabled, to a domain that could have multiple solutions that can be used to fulfill the requirements under certain constraints.

The ability to shift from static security controls that take time to change with unified configuration in most parts of a network, into a more dynamic security capabilities that are temporal (i.e. change with time) and are updated according to the surrounding changes in the environment, will ensure efficient protection, detection and mitigation of attacks.

In this section we highlight the two categories of requirements, and how they provide different security levels.

\subsection{Protection Requirements}
\begin{figure*}[!ht]
	\centering
	\includegraphics[width=.85\linewidth]{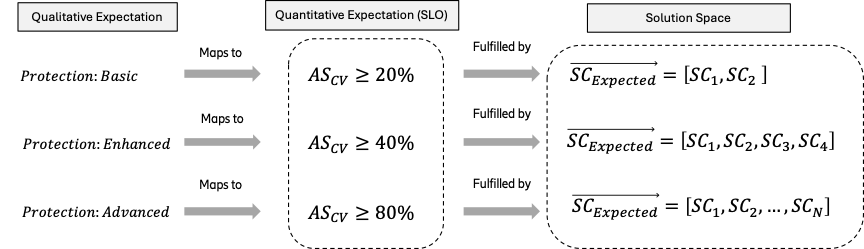}
	\caption{Different levels of protection can be defined qualitatively and achieved through quantitative and measurable expectations}
	\label{fig:protectionLevels}
\end{figure*}

To keep network functions safe from attacks, specific protection features need to be in place. Protection requirements demands implementing such features to protect the areas (called "attack surfaces") where an attacker could try to break in. Functional requirements can be reformulated as goal-driven objectives, resulting in a set of qualitative goals named “Protection Coverage” levels (e.g., basic, enhanced, or advanced protection). These qualitative goals can then be mapped to specific, measurable targets (referred to as "Attack Surface Coverage") that indicate how much of the system is protected. A straightforward approach to this mapping for each "Protection Coverage" level is to define a range for each measurable target that must be met. These ranges serve as Security Level Objectives (SLOs) that define the set of protection features needed to achieve the intended protection goal, as illustrated in Figure \ref{fig:protectionLevels}.

Eventually, the goal of an IbM system is to autonomously infer the appropriate set of security capabilities needed to protect the relevant attack surfaces, fulfilling the qualitative expectation conveyed by the intent. Allowing for different levels of functional requirements is essential, particularly in cloud-based and future mobile networks, where the attack surfaces are expanding and each protection feature has an associated cost. Providing a balance between desired security level and corresponding costs becomes a necessity in the domain of security management.

\subsubsection{Qualitative Expectation}
For the sake of simplicity, in this work we have considered three qualitative "Protection Coverage" levels:
\begin{itemize}
    \item \textbf{ProtectionCoverage-Basic}: This level includes security capabilities that typically enabled by default (e.g., Routing logical separation, logging, authorization and authentication), straightforward to configure, and have minimal impact on computational resource or other KPIs. 
    \item \textbf{ProtectionCoverage-Enhanced}: This level includes security capabilities that are more complex to configure and may have dependencies (e.g., basic firewall rules, role based access control), but do not impose significant computational overhead nor require specialized hardware.
    \item \textbf{ProtectionCoverage-Advanced}: This level includes security capabilities that require complex configuration, are hard to scale, and may have direct or indirect negative impact on available computing resources and overall system performance (e.g., Certifica-based IPsec, User Plane Integrity Protection or hardware-based encryption).
\end{itemize}

\subsubsection{Quantitative Expectation}
The quantitative goal Attack Surface Coverage denoted as ($AS_{CV}$) can be formulated as the ratio of the expected number of attack surfaces that are protected over the total number of attack surfaces existing in a network function. This goal can be simply defined as the below

\begin{equation}
    AS_{CV} = \frac{\sum AS_{Expected}}{\sum AS_{NF}}
    \label{equ:attackSurfaceCoverage}
\end{equation}

Observability is the corner stone and main enabler for intent-based management. It can be seen as the capability to evaluate the level of fulfillment of the quantitative goals expressed by an intent via the observation of the environment. Observability entails collecting data, and calculating a metric that in return is an input to further phases in an automation loop, for example the analysis and planning or evaluation phases as defined in \cite{kephart2003vision}. Thus, the defining $AS_{CV}$ as goal is essential. $AS_{CV}$ is fulfilled by a set of expected security controls that are configured on the network function. Which can be observed by a metric that can be calculated using the below formulation which was defined in \cite{abdelrazek2025measuring}:

\begin{equation}
    SC_{CV} = \frac{\sum SC_{Implemented}}{\sum SC_{Expected}}
    \label{equ:securityControlCoverage}
\end{equation}

\subsubsection{Discussion}
To summarize, we define the \textit{Protection Requirements} as the goals to fulfill the functional security requirement of a NF by implementing and configuring the needed preventative capabilities. Where it can be observability using continuously monitored that the implemented security controls and configuration parameters are the expected to achieve a certain goal, where the metric is defined as Security Control Coverage Metric, or $SC_{CV}$. Finally, differentiation protection levels can be Hybrid meaning that they can have a Qualitative level that is mapped to a Quantitative value, allowing flexibility to be used depending on the type of intent, if its a service intent or an operational intent.

\subsection{Security Performance Requirements}
While functional requirements establish a baseline, defining the concrete security features and functionalities that a system should implement, non-functional requirements refers to how well implemented features performs security-related tasks. \textit{Security Performance Requirements} address aspects such as security capability performance and effectiveness, security state, mean-time-to detect attacks, availability, and others that provides a characteristic of a security posture of the environment or the set of capabilities used to protect, detect and mitigate attacks. 

\begin{figure}[!ht]
	\centering
	\includegraphics[width=0.8\linewidth]{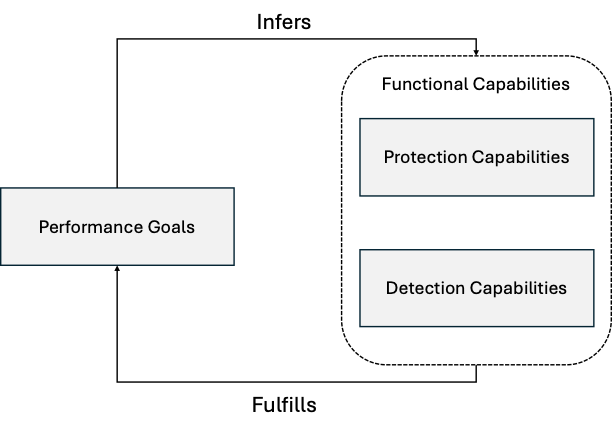}
	\caption{Relationship between Performance Goals and Security Capabilities}
	\label{fig:relation}
\end{figure}

Such requirements aim at providing more observability on the NF security. They are the evolution of the current view on security requirements. They shall become the main goals defined as intents. The non-functional requirements should eventually replace the functional requirements and become the main Security Service Level Objectives (SecSLO) to operate and use as intents. Non-functional requirements infer and deliver the security capabilities required. An example of security performance requirements can be maintaining robustness level at certain conditions of a network function, or ensuring the availability through security capabilities (i.e. ensure suitable controls effectively configured to detect and mitigated denial of service attacks), or ensuring that the needed segmentation in the networks using the suitable controls. The functional capabilities shall be continuously monitored and updated to fulfill the goals and requirements. This relationship between non-functional requirements and functional security capabilities can be illustrated in Figure \ref{fig:relation}.

Performance requirements are related to the performance goal and the security posture. They are quantifiable expectations ensuring that the security controls are performing effectively to achieve the required goals. Continuously evaluate the security posture of the underlying system. 

\subsubsection{Quantitative Expectation}
Similar to the previous case, \textit{Performance Requirements} can be expressed in terms of both qualitative and quantitative expectations. Example of different security levels of performance requirements can be expressed and quantitatively as \textit{Segmentation level} $\ge40\%$ another example could be an expectation for \textit{Radio Robustness level} $\ge60\%$, a third example could be ensuring the detection capabilities are optimized with suitable features and thresholds by setting a goal such as \textit{Mean-time-to-Detect Radio Attacks} $\le200$ms. Performance goals can be in the form of continuous values scale or automation time loops as guiding factors for different tiers (i.e. levels) of performance requirements.

\subsubsection{Discussion}
To summarize, we define the \textit{Performance requirements} as the goals that ensure a NF meets at any time the specified performance goals. These goals may include trusted evaluation of all NF components, the effectiveness of implemented controls, and the threat level calculation. The metrics are dependent on the specific goals set and they provide quantitative measures of performance.

For example, a Threat level metric can be used to evaluate the NF's robustness against potential security threats, while a security control effectiveness metric can assess the fulfillment of segmentation requirements for example, etc... These requirements must be quantitative in nature as they need to provide clear and measurable foundations for assessing the performance of security controls.

\section{Expressing Security Requirements as Intents}~\label{sec_req_intents}
In this section, we attempt to provide a use case to exemplify how the previously mentioned security requirements can be expressed in an intent model. Noting that the intent model is a machine readable model that is translated either from a human natural language or decomposed from another intent. The intent model is relying on ontologies defined by TM Forum mainly the Intent Ontology \cite{tmfimo}, Logical Operators Ontology \cite{tmflog}, Quantity Ontology \cite{tmfquan} and the Security Ontology\cite{tmfsec}. Nevertheless, these models can be mapped to the 3GPP intent modeling scheme using YAML.

\begin{figure}[!ht]
	\centering
	\includegraphics[width=0.95\linewidth]{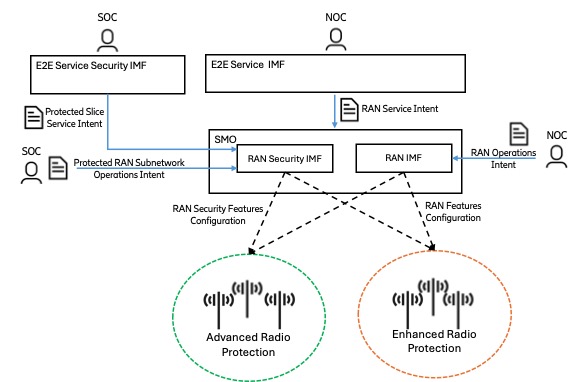}
	\caption{Security Requirements handled by Security Intent Management Function}
	\label{fig:deliveryarch}
\end{figure}

The intent security expectations can be defined differently depending on the type of the intent for example service or operations intent. As a service intent the protection requirements can be defined as qualitative expectations, while on the operations intent the requirements can be defined as quantitative expectations, and mapping between the expectations can occur when a service intent is further decomposed into an operations intent as illustrated in Figure~\ref{fig:deliveryarch}.

The use case presented in this paper exemplifies the delivery of a RAN sub-network with different security expectations, manifested in differentiated protection levels on the available attack surfaces on a network function. The intent expectation to deliver a RAN sub-network with a protection level will be fulfilled by the suitable security capabilities. The presented intent can be formulated as a sub-requirement of delivering an E2E secure slice, where each network domain may have its own security requirement. This intent expectation simplifies the management of deploying and scaling the required security capabilities by:
\begin{itemize}
    \item Reducing time to deploy (i.e. a smaller number of maintenance windows scheduled)
    \item Streamlining and reducing planning efforts
\end{itemize}

The expression of the intent model for this use case, models an operations requirement for the RAN sub-network, where it indicates that there are few base stations that support both generations 4G and 5G located in a specified area would be expected to have an advanced protection level on the air interface, which can be expressed in a human natural language as \textit{Deliver secure gNB and eNB in a location area A with \textbf{advanced protection level on the air interface}} as illustrated in Figure~\ref{fig:deliver_advanced}.

\begin{figure}[!ht]
	\centering
\begin{lstlisting}[language=Turtle]
ex:DeliverSecureRanIntent
    a icm:Intent;
    log:allOf ( ex:DeliverSecureRanNetwork ex:AdvancedRadioProtectionExpectation  );
    rdfs:comment "Intent to deliver an advanced secure Radio interfaces in a RAN subnetwork".
# The assets to be protected
ex:SecureRadio
    a sec:ProtectedInterface;
    sec:interfaceType ex:Uu;
    ex:hasStack ex:ControlPlane, ex:UserPlane;
    sec:hasSecurityProperty sec:Confidentiality, sec:Integrity.
ex:SecureBb01
    a ex:NetworkFunction;
    set:intersection ( [ set:resourcesofType ( ex:gNB ex:eNB ) ] 
                     [ set:resourcesWithPropertyObject ( ex:hasLocationArea "A" ) ] );
    sec:hasProtectedAsset ex:SecureRadio .
# The target object that will fulfill the intent
ex:Target01
    a icm:Target;
    rdfs:member ex:SecureBb01.
# The delivery expectation
ex:DeliverSecureRanNetwork
    a icm:DeliveryExpectation;
    icm:Target ex:Target01. 
# The measurable property expectation
ex:AdvancedRadioProtectionExpectation
    a icm:PropertyExpectation;
    icm:target ex:Target01;
    log:allOf ( ex:AdvancedRadioProtectionCondition ).
# The condition that should be met to fulfill the property expectation
ex:AdvancedRadioProtectionCondition
    a icm:Condition;
    quan:atLeast ( ex:RadioAttackSurfaceCoverage
                    [ rdf:value "0.8"] );
    quan:atMost ( ex:RadioAttackSurfaceCoverage
                    [ rdf:value "1.0"] ).
\end{lstlisting}
	\caption{RAN Advanced Radio Protection Requirements Expressed as a Machine Readable Intent Model using TM Forum Ontologies}
	\label{fig:deliver_advanced}
\end{figure}

While another intent could be modeled as presented in Figure~\ref{fig:deliver_enhanced}, where a set of 5G base stations in a different area they only require enhanced protection level provided as input as \textit{Deliver secure gNB in a location area B with \textbf{enhanced protection level on the air interface}}. Enhanced protection on radio interface means providing the suitable security features to ensure confidentiality and integrity properties on the radio control plane. On the other hand, an advanced protection level would result in providing protection not only on control plane but also for the radio user plane.

\begin{figure}[!ht]
	\centering
\begin{lstlisting}[language=Turtle]
ex:DeliverSecureRanIntent
    a icm:Intent;
    log:allOf ( ex:DeliverSecureRanNetwork ex:EnhancedRadioProtectionExpectation );
    rdfs:comment "Intent to deliver an enhanced secure Radio interfaces in a RAN subnetwork".
# The assets to be protected
ex:SecureRadio
    a sec:ProtectedInterface;
    sec:interfaceType ex:Uu;
    ex:hasStack ex:ControlPlane;
    sec:hasSecurityProperty sec:Confidentiality, sec:Integrity.
ex:SecureBb02
    a ex:NetworkFunction;
    set:intersection ( [ set:resourcesofType ( ex:gNB ) ] 
                     [ set:resourcesWithPropertyObject ( ex:hasLocationArea "5678" ) ] );
    sec:hasProtectedAsset ex:SecureRadio .
# The target object that will fulfill the intent
ex:Target02
    a icm:Target;
    rdfs:member ex:SecureBb02.
# The delivery expectation
ex:DeliverSecureRanNetwork
    a icm:DeliveryExpectation;
    icm:Target ex:Target02. 
# The measurable property expectation
ex:EnhancedRadioProtectionExpectation
    a icm:PropertyExpectation;
    icm:target ex:Target02;
    log:allOf ( ex:EnhancedRadioProtectionCondition ).
# The condition that should be met to fulfill the property expectation
ex:EnhancedRadioProtectionCondition
    a icm:Condition;
    quan:atLeast ( ex:RadioAttackSurfaceCoverage
                    [ rdf:value "0.5"] );
    quan:atMost ( ex:RadioAttackSurfaceCoverage
                    [ rdf:value "0.79"] ).
\end{lstlisting}
	\caption{RAN Enhanced Radio Protection Requirements Expressed as a Machine Readable Intent Model using TM Forum Ontologies}
	\label{fig:deliver_enhanced}
\end{figure}

\section{Standardization Discussion} \label{standards}

By leveraging on security requirements as standardized intents we aim at improving interoperability and, most importantly, allowing for the definition of differentiated protection level tailored to the specific needs of NFs and use cases. 
This section examines the current landscape of relevant standards, identifies key gaps, and proposes how the present work can contribute to ongoing standardization efforts in the security management domain.

As the expectations and requirements expressed by intents are agnostic to system implementation and technology for their realization, standardization of security intents provides a first layer of interoperability for autonomous security management in mobile network operators environment. As previously discussed, automating security management is considered a fundamental step for reducing OPEX and support new business models (e.g., providing network slices with differentiated security capabilities)

Notable standard bodies have established normative initiatives for IbM. In 3GPP, SA5 Working Group is responsible for the specification of intent management features that enable operators to express standardized intent expectation and goals, where intents are used to express high-level service requirements and network behaviors~\cite{3ggp28312}. TMF, as part of the autonomous network framework, has standardized intent ontology and related frameworks supporting intent-driven operations~\cite{IG1253,tmfimo}. A first example of security intent ontology targeting mobile network domains can be found in~\cite{tmfsec}. The IRTF Network Management Research Group has proposed an intent classification methodology and intent taxonomy~\cite{rfc9316}. There, security characteristics are considered in the intent scope for both networks and service intents. 
On the other hand, current standards do not yet provide formal definitions for either qualitative and quantitative expectations and metrics for the security domain that could be expressed as intents. This absence hinders today the ability of network operators to define and implement differentiated security objective levels tailored to the needs of deployed network services and infrastructure. Consequently, security management remains largely configuration-driven, which lacks adaptability, flexibility and support for differentiation of security. This gap is particularly concerning for highly dynamic environments, where security capabilities need to be adapted in real-time to match a changing threat landscape. 

By building on and extending existing standardization efforts, the proposed approach enable to specify intent-based SSLA by introducing both Protection and Performance Requirements as set of expectations. In Section~\ref{sec_req_intents} it is shown an examples of the feasibility of expressing intent models using the RDF language. As per, Figure~\ref{fig:deliveryarch}, such intents can be applied both at the service and operations layer. 
Some levels of security observability is another direction that should be pursued in standardization to allow multi-domain security operation.  

Standardization of such, allows the different vendors and platform to interoperate, an increasingly important factor for service providers and network operators managing multi-domain, multi-vendor, and multi-tenant 5G/6G systems. By expressing unambiguous security requirements and goals via intents in a standardized way, the proposed approach offers several key industry benefits: 
\begin{itemize}
    \item Network vendors maintain the flexibility to implement and fulfill security intents using the techniques and technologies of their choice. This will foster innovation and competition in providing both efficient and effective security solutions;
    \item A standardize intent model supports high-level of interoperability and low-coupling between the various components of mobile networks, minimizing the risk of vendor lock-in;
    \item Network operators can start differentiating their offerings based on advanced security capabilities. This will concur in promoting competition and deliver greater value to end customers. 
\end{itemize}

Other expected benefits of standard-compliant, interoperable security management solutions include reduction in OPEX, and improved scalability, as new nodes and subsystems can be integrated more efficiently.

\section{Conclusion}\label{conclusion}
In this paper, we have highlighted that current security management in mobile networks is lacking  simple, unambiguous methods for configuring and operating system characteristics. This limitation not only hinders automation, but ultimately undermines the overall security posture of the system in the evolving and varying threat landscape. To address these challenges, we proposed extending the Intent-based Management (IbM) approach to  define and manage different security levels via intents. Such formalization demonstrates that both functional and non-functional security requirements can be expressed and conveyed via intents. We also discussed the need for standardizing these security intents to ensure interoperability. Overall, our findings highlights the potential benefit of adopting intent-based security management for simplifying security operations and ensuring efficient management of security capabilities and requirements in 5G and future networks.

\bibliographystyle{IEEEtran}
\bibliography{bib/references}
\vspace{10pt}

\end{document}